\documentclass[aps,prd,twocolumn,groupedaddress,showpacs]{revtex4-1}

\usepackage{graphicx}% Include figure files
\usepackage[caption = false]{subfig}

\usepackage{amsmath}
\usepackage{amssymb}

\begin{document}

%\preprint{}

\title{``Big Bang'' as a result of first-order phase transition driven by changing scalar curvature in expanding early Universe: ``hyperinflation'' scenario}


\author{E.A. Pashitskii}
\email[]{pashitsk@iop.kiev.ua}
%\thanks{}
\affiliation{Institute of Physics, NAS of Ukraine, 46 Nauki Avenue, Kiev, 03028, Ukraine}

\author{V.I. Pentegov}
\email[]{pentegov@iop.kiev.ua}
%\thanks{}
\affiliation{Institute of Physics, NAS of Ukraine, 46 Nauki Avenue, Kiev, 03028, Ukraine}

%\date{\today}
\date{December 29, 2014}

\begin{abstract}
We suggest that the ``Big Bang'' may be a result of the first-order phase transition driven by changing scalar curvature of the 4D space-time in expanding cold Universe, filled with nonlinear scalar field $\phi $ and neutral matter with equation of state $p=\nu \varepsilon $ (where $p$ and $\varepsilon $ are pressure and energy density of matter). We consider a Lagrangian for scalar field in curved space-time with nonlinearity $\phi ^{4} $, which along with the quadratic term $-\xi R\left|\phi \right|^{2} $ (where $\xi $ is the interaction constant and $R$ is scalar curvature) contains a term $\sim \xi R\left(\phi +\phi ^{+} \right)$ linear in $\phi $. Due to this term the condition for the extrema of the potential energy of scalar field is given by a cubic equation. Provided $\nu >1/3$ the scalar curvature $R=[\kappa (3\nu -1)\varepsilon -4\Lambda ]$ (where $\kappa $ and $\Lambda $ are Einstein's gravitational and cosmological constants) decreases along with decreasing $\varepsilon $ in the process of the Universe's expansion, and at some critical value $R_{c} <0$ a first-order phase transition occurs, induced by an ``external field'' parameter proportional to $R$. Given certain conditions the critical radius of the early Universe at the point of the first-order phase transition may reach arbitrary large values, so this scenario of unrestricted ``inflation'' of the Universe may be called ``hyperinflation''. Beyond the point of phase transition the system is rolling down into the potential minimum with the release of the potential energy of scalar field, accompanied by oscillations of its amplitude $\phi $. The powerful heating of the Universe in the course of attenuation of these oscillations plays the role of ``Big Bang''.
\end{abstract}

\pacs{98.80.Bp, 98.80.Cq}
%\keywords{}

\maketitle


\section{Introduction}

The question about the cause of the ``Big Bang'' starting the hot phase in the development of our Universe is still a central question of cosmology. The earliest stage of evolution of the cold pre-``Big Bang'' Universe, which now may be traced only through the manifestations of the relict gravitational radiation, was considered first in the works of Gliner~\cite{Gliner1969,Gliner1970} and Starobinsky~\cite{Starobinskii1979,Starobinsky1980}.

In the same time, starting with the works of Kirzhnits and Linde~\cite{Kirzhnits1972,Kirzhnits1972a,Kirzhnits1975,Kirzhnits1976} and Guth~\cite{Guth1981} (see also~\cite{Kazanas1980,Sato1981,Albrecht1982}), various inflation scenarios of the initially hot Universe were investigated. In these scenarios the first- or second-order phase transitions induced by temperature occurred in the process of expansion and cooling of the Universe with subsequent spontaneous braking of symmetries of various interactions and production of fields and particles from vacuum. A number of authors also considered first-order phase transitions, driven by the density changes~\cite{Harrington1974,Krive1976,Lee1974,Linde1976a,Linde1979,Krive1982} or by external fields and currents~\cite{Salam1974,Salam1975,Linde1976,Krive1976a}. The common shortcoming of the hot Universe scenarios is the existence of critical fluctuations during the second-order phase transitions or the formation of domains (``bubbles'') of new phase in the case of the first-order phase transitions, leading to a strong large-scale spatial inhomogeneity of matter and anisotropy of the relic radiation in the modern Universe, which contradicts astronomical observations.

Thereby Linde~\cite{Linde1982,Linde1983,Linde1983a,Linde1990} proposed a scenario of ``chaotic inflation'' for the early cold Universe, when the energy density of vacuum was determined by the potential energy density of a nonlinear scalar field $U\left(\phi \right)\le M_{P}^{4} $ (where $M_{P} \sim {1\mathord{\left/ {\vphantom {1 \sqrt{G} }} \right. \kern-\nulldelimiterspace} \sqrt{G} } $ is the Planck mass, presuming $\hbar =c=1$, and $G$ is the Newton gravitational constant). For the potentials $U\left(\phi \right)$ described by powers of $\phi $ with sufficiently small interaction constants and large initial values of the field amplitude $\phi \gg M_{P} $, the initial quantum fluctuations on the Planck scale $l_{P} \sim 1/M_{P} $ expand (``inflate'') gigantically, by many orders of magnitude exceeding the observable size of the present-day Universe, which provides explanation for its flat geometry, isotropy and high degree of large-scale homogeneity, as well as for the absence of domain walls and t'Hooft~\cite{tHooft1974} -- Polyakov~\cite{Polyakov1974} monopoles in our world.

In the inflation scenarios the heating of the cold Universe to high temperatures (so called ``reheating'', actually playing the role of ``Big Bang'') occurs as the result of dissipation of the energy of oscillations of the scalar field's amplitude, with energies of about $10^{12} -10^{14}$~GeV, in the vicinity of the potential minimum. The attenuation of these oscillations is due to both the expansion of the Universe and the production of various particles and antiparticles~\cite{Kofman1997}.

Later on a scenario of ``hybrid inflation''~\cite{Linde1994} was put forward, according to which there existed two different types of scalar fields in the early Universe, with significantly different equilibrium amplitudes and velocities of rolling down into the potential minimum. This approach allowed to reconcile the inflation theory with the theory of supergravitation~\cite{Linde2004}.

It is well to bear in mind though, that given the small scale and high scalar curvature of the early Universe the interaction of the fundamental scalar field with gravitation should have played a significant role in the processes of its evolution, which was not considered in~\cite{Linde1982,Linde1983,Linde1983a,Linde1990,Kofman1997}. In accordance with the general principles of the quantum field theory in the 4-space with finite scalar curvature $R\ne 0$ (see~\cite{Birrell1982}), the initial Lagrangian of the nonlinear scalar field $\phi $ should contain a term ${R\left|\phi \right|^{2} \mathord{\left/ {\vphantom {R\left|\phi \right|^{2}  6}} \right. \kern-\nulldelimiterspace} 6} $~\cite{Krive1976} (see also~\cite{Spokoinyi1984}) quadratic in $\phi $, where the coefficient ${1\mathord{\left/ {\vphantom {1 6}} \right. \kern-\nulldelimiterspace} 6} $ ensures the conformal invariance of the theory in the limit of zero bosonic mass $\mu \to 0$. This leads to the renormalization of the parameter of self-action for the scalar field and of the Einstein gravitational constant $\kappa =8\pi G$ in the general relativity equations by the factor of about $\kappa \left|\phi \right|^{2} /3$. In particular, for the Higgs field~\cite{Higgs1964,Higgs1964a} with vacuum average $\phi _{H} \equiv \upsilon \approx 247$~GeV (see~\cite{Weinberg1996}) this renormalization is anomalously small and has the order of magnitude of $G/G_{F} \sim 10^{-32} $ (where $G_{F} $ is the Fermi constant for the weak interaction).

As was shown in~\cite{Zee1979,Smolin1979,Cervantes-Cota1995}, the term of a more general form $-\xi R\left|\phi \right|^{2} $ in the Lagrangian of the Higgs field with $\mu \ne 0$, where the dimensionless constant $\xi $ may be treated as the constant of interaction of scalar and gravitational fields, leads in the framework of standard model to generation of mass of the order of the Planck one $\left(M_{P} \right) $ only for anomalously large values of the constant $\xi \ge 10^{34} $.

Another expression for this constant $\xi \approx 4\cdot 10^{4} \cdot {m_{H} \mathord{\left/ {\vphantom {m_{H}  \upsilon \sqrt{2} }} \right. \kern-\nulldelimiterspace} \upsilon \sqrt{2} } $, where $m_{H} $ is the mass of the Higgs boson, was obtained by the authors of~\cite{Bezrukov2008} for some modified exponentially flat potential of the nonlinear scalar field in the regime of ``slow roll'' of the system into the ground state. Bearing in mind the recently established value $m_{H} \approx 125.5$ ~GeV~\cite{ATLASCollaboration2012,CMSCollaboration2012} we obtain the magnitude of for the constant of interaction of the Higgs field with gravitation $\xi \approx 1.44\cdot 10^{4} $, which is still quite large.\textbf{}

In the present paper, contrary to the Guth~\cite{Guth1981} scenario with the first-order temperature-driven phase transition in the expanding Universe, we propose an alternative scenario of the evolution of the early cold Universe with the first-order phase transition, induced by the parameter of an ``external field'' proportional to scalar curvature. It is shown that this transition is possible given the following conditions:

(i) the Universe born in a rather large quantum fluctuation is filled with some fundamental scalar field $\phi $, described by the Lagrangian which in the curved 4-space with $R\ne 0$ includes a term $\sim \xi R\left(\phi +\phi ^{+} \right)$ linear in $\phi $, playing the role of an ``external field'', along with the quadratic in $\phi $ term $-\xi R\left|\phi \right|^{2} $.

(ii) the early Universe with finite cosmological constant $\Lambda $ contains neutral cold matter, described by the equation of state $p=\nu \varepsilon $ with $\nu >1/3$, which ensures the decreasing of the scalar curvature $R=[\kappa (3\nu -1)\varepsilon -4\Lambda ]$ along with the decreasing of the matter's energy density in the process of the Universe expansion.

In a sense, the model of evolution of the early cold Universe proposed hereafter may be viewed as a modification of the model of ``hybrid inflation''~\cite{Linde1994}, where the role of the second (auxiliary) field is played by the ``external field'' parameter.

In section~\ref{Sec_Lagrangian} of this paper we consider a modified Lagrangian of some fundamental complex scalar field $\phi $ with nonlinearity of $\phi ^{4} $ type, interacting with gravitational field, which is described by the term $-\xi R\left|(\phi -\phi _{0} )\right|^{2} $, where $\phi _{0} $ is the vacuum average of the scalar field amplitude. Hence, the Lagrangian contains both the standard term $-\xi R\left|\phi \right|^{2} $ quadratic in $\phi $ and the term $\xi R\phi _{0} \left(\phi +\phi ^{+} \right)$ linear in $\phi $ and $R$. The equation determining the extrema of the scalar field's potential $U\left(\phi ,R\right)$ is a cubic one with respect to the real part of $\phi $, having three real roots for a certain range of $R$ values, describing two minima and one maximum of the potential $U(\phi ,R)$.

In section~\ref{Sec_Transition} we introduce the dimensionless variables and obtain the potential of the nonlinear scalar field as the function of its amplitude for various values of the dimensionless ``external field'' $h={2\xi R\mathord{\left/ {\vphantom {2\xi R \mu ^{2} }} \right. \kern-\nulldelimiterspace} \mu ^{2} } $. We plot the $h$ dependencies of the real roots of the cubic equation and extremal values of the potential, and use this dependencies to analyze the conditions for the first-order phase transition in metastable state.

In section~\ref{Sec_Evolution} we describe the parameter domains where self-consistent solutions exist for the general relativity equations describing the expanding early cold Universe, filled with scalar nonlinear field and neutral matter with equation of state $p={2\varepsilon \mathord{\left/ {\vphantom {2\varepsilon  3}} \right. \kern-\nulldelimiterspace} 3} $, which corresponds to a neutral non-relativistic ideal gas of massive fermions (see~\cite{Landau1980}). It is assumed that this Fermi-gas consists of the equal numbers of particles and antiparticles with half-integer spin, which are born from the vacuum as the result of a large quantum fluctuation and obtain finite mass due to the interaction with scalar field, in accordance with the Higgs mechanism of mass generation~\cite{Higgs1964,Higgs1964a}. The interaction between fermions is taken to be weak, so that the time of particle-antiparticle annihilation significantly exceeds the time of the early Universe's evolution to the point of phase transition. For  nonzero energy density of the vacuum $\lambda $, which is determined by the energy density of the scalar field in the potential minimum where the early Universe resides, we obtain numerical solutions of the nonlinear general relativity equations, which give for various parameters the time dependencies of the Universe's radius up to the moment of phase transition $t_{c} $, when the radius reaches maximal value $a_{c} =a(t_{c} )$. We show that these solutions exist only for large enough initial values of the radius of quantum fluctuation $a_{0} \ge 4.5l_{P} $, while the radius $a_{c} $ of the early Universe in the transition point and the scalar field's energy $E_{c} \sim a_{c}^{3} $, released in the phase transition, diverge with $\xi \to \xi _{\min } $, where $\xi _{\min } $ is some limiting value of $\xi $, dependent on the parameters of the scalar field. Such a regime of unbounded inflation of the early Universe may be called ``hyperinflation''. The time of the Universe's evolution $t_{c} $ diverges as well when $\xi \to \xi _{\min } $ and the initial radius $a_{0} $ of quantum fluctuation approaches some limiting value $a_{0 \min } (\xi )$. One should remember though that $t_{c} $ may not exceed the annihilation time of the fermions and antifermions in the early Universe.

In section~\ref{Sec_Model} we discuss possible values of the parameters $\mu $ and $g$ of the fundamental nonlinear scalar field, which define the vacuum average $\phi _{0} =\mu /g$ and the potential energy density $U\sim \mu ^{2} \phi _{0}^{2} $. We show that in the case of $\phi _{0} \gg \phi _{H} $ the constant $\xi \ll 1$, contrary to those models where the fundamental field is assumed to be the Higgs field with dimensionless parameter $\kappa \phi _{H}^{2} \approx 10^{-32} $, thus giving $\xi \ge 10^{30} $ (see~\cite{Zee1979,Smolin1979,Cervantes-Cota1995}). In particular, for the ratio $\phi _{0} /\phi _{H} \approx 10^{16} $, when $\kappa \phi _{0}^{2} \approx 1$, the constant $\xi $ is limited from below by the minimal critical value $\xi _{\min } \approx 0.04$ and the value of $\xi $ should be close to $\xi _{\min } $ for the Universe to expand to a significant size of $a_{c} \gg a_{0} \gg l_{P} $. We also estimate the frequency of oscillations of the scalar field amplitude, which are attenuated due to both the Universe's expansion and the birth of a large number of various particle-antiparticle pairs from vacuum (see e.g.~\cite{Kofman1997}). A rapid heating of the Universe, playing the role of ``Big Bang'', should occur due to the large energy of scalar field $E_{c} =\Delta U\cdot \upsilon _{c} $ released in the first-order phase transition in the volume of the closed Universe $\upsilon _{c} =2\pi ^{2} a_{c}^{3} $.

\section{\label{Sec_Lagrangian} Modified Lagrangian of a nonlinear scalar field in curved space-time}

The Lagrangian of a complex scalar field with nonlinearity $\phi ^{4} $ and imaginary ``mass'' $i\mu $ in the curved 4-space with metric tensor $g^{\mu \nu } $ and finite scalar curvature $R\ne 0$, satisfying the condition of conformal invariance in the limit $\mu \to 0$, is written as~\cite{Krive1976}
\begin{multline} \label{EQ_1}
    L=g^{\mu \nu } \left(\partial _{\mu } \phi \right)\left(\partial _{\nu } \phi ^{+} \right)+\mu ^{2} \phi \phi ^{+} -g^{2} \left(\phi \phi ^{+} \right)^{2} \\ +{R\phi \phi ^{+} \mathord{\left/ {\vphantom {R\phi \phi ^{+}  6}} \right. \kern-\nulldelimiterspace} 6},
\end{multline}
where $g^{2} $ is the parameter of nonlinearity (self-action) of the scalar field. The last term in Lagrangian \eqref{EQ_1}, as shown in~\cite{Krive1976}, leads to the renormalization of the constants $\mu ^{2} $ and $g^{2} $, as well as the Einstein gravitational constant $\kappa =8\pi G$ in the equations of general relativity, by the dimensionless quantity $\kappa \phi _{0}^{2} $. As was mentioned earlier, for the Higgs field this renormalization is quite small (of the order of $G/G_{F} \sim 10^{-32} $)/

In the more general form with $\mu \ne 0$ the Lagrangian \eqref{EQ_1} may be written as (see~\cite{Bezrukov2008}):
\begin{equation} \label{EQ_2}
    L=g^{\mu \nu } \left(\partial _{\mu } \phi \right)\left(\partial _{\nu } \phi ^{+} \right)+\frac{\mu ^{2} }{2} \left|\phi \right|^{2} -\frac{g^{2} }{4} \left|\phi \right|^{4} -\xi R\left|\phi \right|^{2},
\end{equation}
where $\xi $ is the effective dimensionless constant of interaction between scalar and gravitational fields. As we can see, for nonzero curvature of the 4-space the parameter $\mu ^{2} $ is renormalized as $\mu ^{2} \to \left(\mu ^{2} -2\xi R\right)$, and there is still a possibility of the second-order phase transition with the curvature-dependent order parameter $\phi _{0} (R)=\sqrt{(\mu ^{2} -2\xi R)} /g$. The mass of the scalar boson, analogues to the Higgs boson, is written in this case as $m_{B} (R)=\sqrt{2(\mu ^{2} -2\xi R)} $.

In the present paper we consider a certain fundamental scalar field $\phi $ with nonlinearity $\phi ^{4} $ and with a modified Lagrangian in the curved space-time, which, along with the term $-\xi R\left|\phi \right|^{2} $ quadratic in $\phi $, contains also a term linear in $\phi $ and $R$. Earlier, in~\cite{Pashitskii2014}, this additional term was chosen in the form of $\zeta R\phi /\sqrt{\kappa } $, where $\zeta $ was some dimensionless constant, not equal to $\xi $ in general case, while the factor $1/\sqrt{\kappa } $ was introduced on the basis of dimensionality consideration, as the dimensionalities of $\phi ^{2} $ and $\kappa ^{-1} $ coincide. However, the introduction of an additional parameter $\zeta \ne \xi $ seems unnecessary.

In the model considered henceforth the interaction of the complex scalar nonlinear field with gravitation is given as $-\xi R\left|(\phi -\phi _{0} )\right|^{2} $, where $\phi _{0} $ is the vacuum average of the scalar field.

As the result, the $\phi ^{4} $ Lagrangian of the scalar field contains linear and quadratic in $\phi $ terms, which are proportional to scalar curvature $R$:
\begin{align} \label{EQ_3}
    \tilde{L}=\frac{1}{2} g^{\mu \nu } \left(\partial _{\mu } \phi \right)\left(\partial _{\nu } \phi ^{+} \right) &+\frac{1}{2} \left(\mu ^{2} -2\xi R\right)\left|\phi \right|^{2} -\frac{1}{4} g^{2} \left|\phi \right|^{4} \nonumber \\ &+\xi R\phi _{0} (\phi +\phi ^{+} )-\xi R\phi _{0}^{2}.
\end{align}

Assuming $\phi =\left(\Phi +\phi '\right)$, where $\Phi (R)$ is the real (classical) part of the field amplitude $\phi $ for $R\ne 0$ and $\phi '$ is its complex (quantum) part, and varying \eqref{EQ_3} with respect to $\phi '$ we obtain in the linear approximation $\left|\phi '\right|\ll \Phi $ the equation for the bosonic field $\phi '$ with the curvature-dependent mass of the scalar boson:
\begin{equation} \label{EQ_4}
    m_{B} (R)=\sqrt{3g^{2} \Phi ^{2} \left(R\right)-\left(\mu ^{2} -2\xi R\right)}.
\end{equation}

In the zero-order approximation in $\phi '$ \eqref{EQ_3} gives the expression for the potential energy density of the real part of the scalar field $\Phi $:
\begin{multline} \label{EQ_5}
    U(\Phi ,R)=  \frac{1}{4} g^{2} \Phi ^{4} -\frac{1}{2} (\mu ^{2} -2\xi R)\Phi ^{2} \\ -2\xi R\phi _{0} (\Phi -\phi _{0} /2)+U_{0} ,
\end{multline}
where $U_{0} $ is some arbitrary constant, which should ensure the zero minimal value of the potential \eqref{EQ_5}.

The condition of the existence of the extrema of potential \eqref{EQ_5} is given by the cubic equation with respect to amplitude $\Phi $:
\begin{equation} \label{EQ_6}
    \frac{\partial U}{\partial \Phi } =g^{2} \Phi ^{3} -\left(\mu ^{2} -2\xi R\right)\Phi -2\xi R\phi _{0} =0.
\end{equation}

As will be shown below, in a certain parametric domain equation \eqref{EQ_6} has three real roots, so the change in the scalar curvature may lead to the first-order phase transition.

\section{\label{Sec_Transition} First-order phase transition in the early Universe with changing scalar curvature}

For the farther analysis of equations \eqref{EQ_5} and \eqref{EQ_6} it is convenient to introduce dimensionless variables $x={\Phi \mathord{\left/ {\vphantom {\Phi  \phi _{0} }} \right. \kern-\nulldelimiterspace} \phi _{0} } $ and $V={U\mathord{\left/ {\vphantom {U \mu ^{2} \phi _{0}^{2} }} \right. \kern-\nulldelimiterspace} \mu ^{2} \phi _{0}^{2} } $:
\begin{equation} \label{EQ_7}
    V\left(x,h\right)=\frac{x^{4} }{4} -\left(1-h\right)\frac{x^{2} }{2} -hx+\frac{h}{2} +V_{0};
\end{equation}
\begin{equation} \label{EQ_8}
    \frac{\partial V}{\partial x} =x^{3} -\left(1-h\right)x-h=0,
\end{equation}
where $h=2\xi R/\mu ^{2} $ is the dimensionless ``external field'' and $V_{0} ={U_{0} \mathord{\left/ {\vphantom {U_{0}  \mu ^{2} \phi _{0}^{2} }} \right. \kern-\nulldelimiterspace} \mu ^{2} \phi _{0}^{2} } $.

\begin{figure}[h]
  \includegraphics[width=\columnwidth]{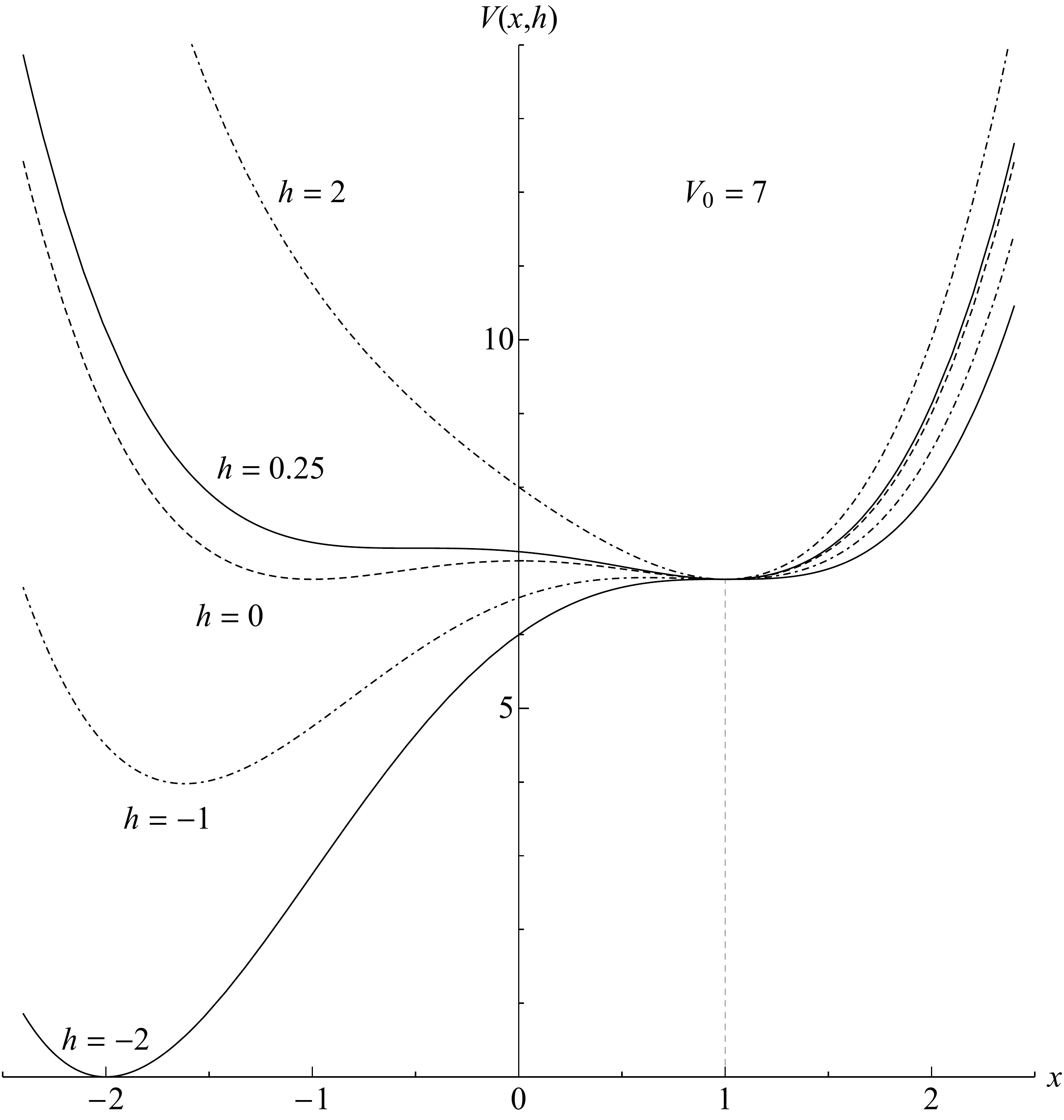}
  \caption{\label{Fig1} The dimensionless potential of the nonlinear scalar field $V\left(x,h\right)={U\left(\Phi ,R\right) / \mu ^{2} \phi _{0}^{2} } $ in function of the dimensionless amplitude $x={\Phi / \phi _{0} } $ for various values of the ``external field'' $h={2\xi R / \mu ^{2} } $. Solid lines represent $V\left(x,h\right)$ for the threshold value $h=0.25$, when the second minimum of the potential appears, and also for the critical value $h=h_{c} \equiv -2$, when one of the minimums of the potential $V\left(x,-2\right)$ merges with the maximum, giving an inflection point at $x= 1$, while the other minimum at $x=-2$ becomes zero for $V_{0} =7$.}
\end{figure}

In Fig.~\ref{Fig1} the $x$ dependencies of the potential \eqref{EQ_7} are shown for various values of the parameter $h$ for $V_{0} =7$. In the region $h\ge 0.25$ potential \eqref{EQ_7} has only one minimum $V_{\min } =6.75$ at the point $x=1$, where the system (the early Universe) resides initially. As we can see, the depth and position of this minimum remain the same for all values of $h$ in the range $-2<h<\infty $. For the ``external field'' $h<0.25$ there appears a second minimum in the potential \eqref{EQ_7}, separated from the first one by a potential barrier. At $h=0$ both minima have equal depth $V_{\min }^{\left(1\right)} =V_{\min }^{\left(2\right)} =6.75$ and positioned symmetrically in points $x=\pm 1$, while the maximum is at the point $x=0$ with $V_{\max } =7$. In the region $h<0$ the left minimum grows deeper with the decreasing parameter $h$. Finally, for $h=-2$ the maximum of the potential \eqref{EQ_7} and its right minimum disappear, giving an inflection point at $x=1$, while the left minimum reaches zero value at the point $x=-2$.

\begin{figure}[h]
  \subfloat[]{\label{Fig2a} \includegraphics[width=.98\columnwidth]{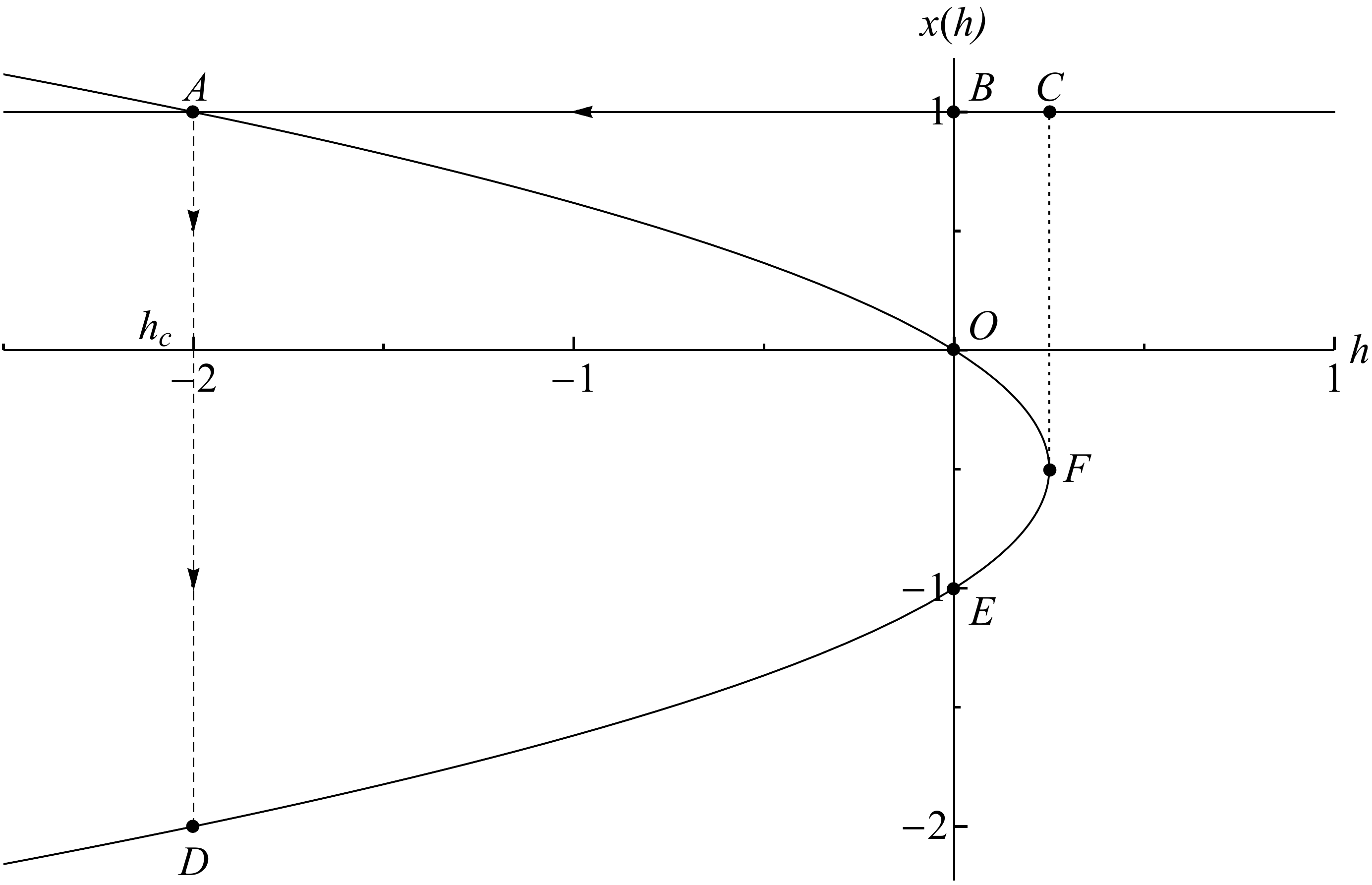}}\\
  \subfloat[]{\label{Fig2b} \includegraphics[width=.98\columnwidth]{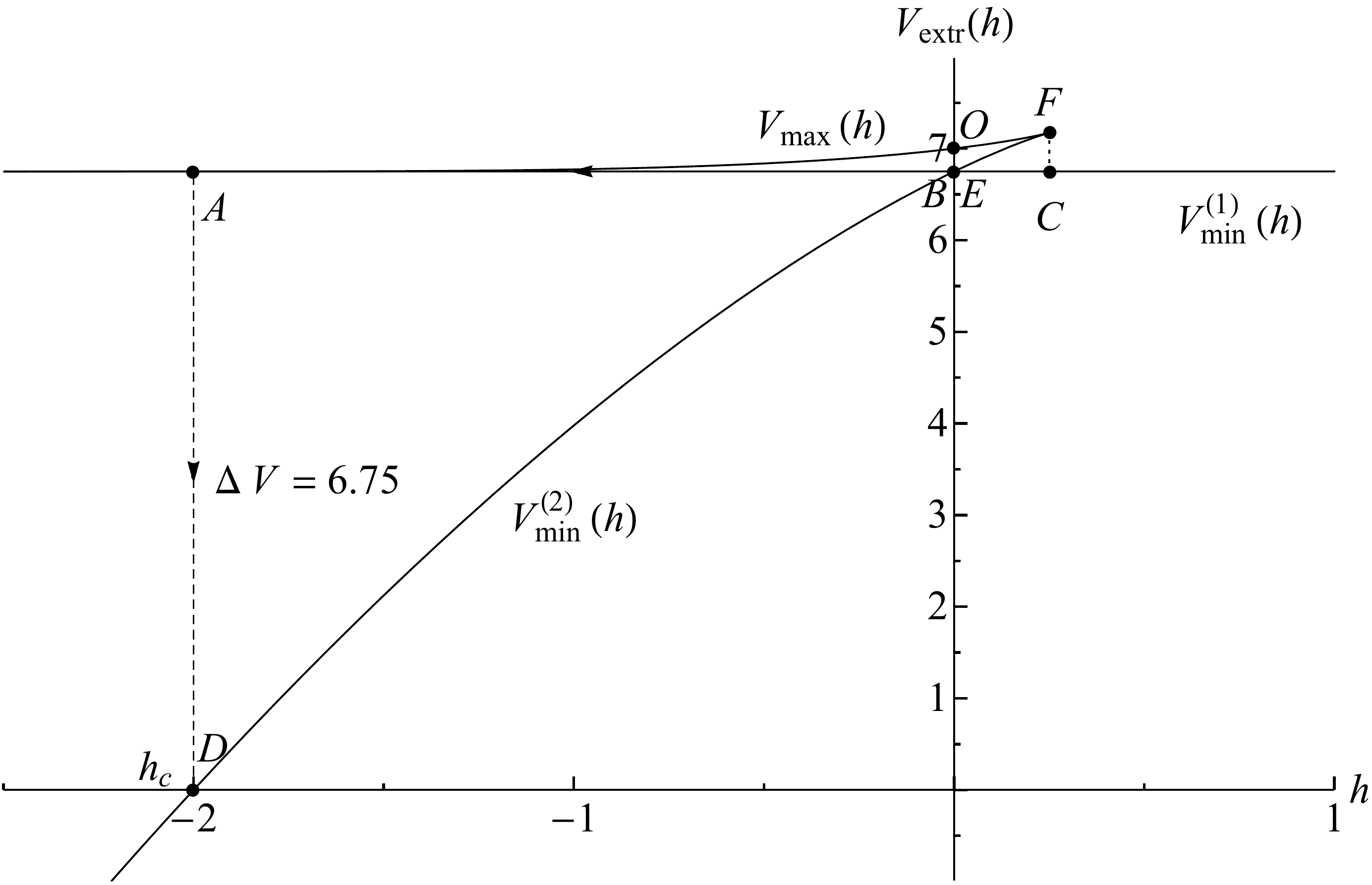}}
  \caption[]{\label{Fig2} The $h$ dependencies of three real roots $x_{i} \left(h\right)$ of the cubic equation \eqref{EQ_8} in the range $-2\le h\le 0.25$ \subref{Fig2a} and corresponding minimal $V_{\min }^{(1,2)} \left(h\right)$ and maximal $V_{\max} \left(h\right)$ values of the potential ${V_{\operatorname{extr} }}\left( h \right) = V\left( {{x_i}\left( h \right),h} \right)$ in the same region \subref{Fig2b}.}
\end{figure}

The tree real roots $x_{i} \left(h\right)$ of the cubic equation \eqref{EQ_8} are shown in Fig.~\ref{Fig2} for $-2\le h\le 0.25$. For all $h>-2$ there is a positive $h$-independent root $x=1$, which corresponds to the value $\Phi =\phi _{0} \equiv \mu /g$ of the amplitude of the scalar field (the $ABC$ branch). This positive root determines the constant position of the right minimum of the potential \eqref{EQ_7}, the negative root (the $DEF$ branch) gives the position of the left minimum, while the sign-changing root (the $AOF$ branch) describes the position of the potential maximum, i.e. it corresponds to the absolutely unstable state. The segments $AB$ and $EF$ on the positive and negative branches correspond to metastable states. Notice that according to \eqref{EQ_4} the boson mass for $\Phi =\phi _{0} $ equals
\begin{equation} \label{EQ_9}
  m_{B} \left(R\right)=\sqrt{2\left(\mu ^{2} +\xi R\right)} \equiv \mu \sqrt{2+h},
\end{equation}
and the value $m_{B} (R)$ should be bounded by the Planck mass $M_{P} $ from above.

Three extremal values of the potential \eqref{EQ_7}, shown in Fig.~\ref{Fig2b}, two minima $V_{\min }^{\left(1\right)} (h)$ and $V_{\min }^{\left(2\right)} (h)$ and one maximum $V_{\max } (h)$, correspond to the three roots of the cubic equation \eqref{EQ_8} of the Fig.~\ref{Fig2a}. In the absence of the nucleation centers of the new phase, when no transitions, accompanied by the formation of domains with alternating sign of the order parameter, occur between the branches $ABC$ and $DEF$ for $-2<h<0.25$, the diminishing of scalar curvature $R$ implies that the system travels along the straight phase trajectory $ABC$ with constant minimal value of the potential $V_{\min }^{\left(1\right)} =6.75$ (for $V_{0} =7$) from the stable state at some initial point with $h>0.25$ into the critical point $A$ with $h=-2$. After that the system drops from point $A$ to point $D$, which corresponds to a phase transition with the decrease in the potential energy density of scalar field by $\Delta V=6.75$.

\section{\label{Sec_Evolution} Evolution of the early cold Universe towards the first-order phase transition}

Suppose that in homogeneous space filled with nonlinear scalar field in ground state at zero temperature with equilibrium amplitude $\phi _{0} $ and minimal potential energy density $U_{\min } (\phi _{0} )=6.75\mu ^{2} \phi _{0}^{2} $ at some (initial) moment appears a rather large quantum fluctuation, spontaneously producing some matter, which is neutral with respect to all charges and characterized by the equation of state $p=\nu \varepsilon $, where $p$ and $\varepsilon $ are pressure and energy density of the matter, with dimensionless coefficient $\nu $ satisfying condition $\nu >1/3$. The initial scalar curvature of the 4-space equals $R_{i} =-4\tilde{\kappa }\lambda $ (where $\tilde{\kappa }$ is the renormalized Einstein constant, $\lambda $ is the energy density of vacuum), while after the emergence of matter it increases to positive values (see below).

Suppose also that due to space isotropy the form of the quantum fluctuation is close to spherical, while its initial radius considerably exceeds the Planck length $l_{P} $, so the description of the evolution of the spherically symmetric incipient Universe may neglect quantum effects, such as the tunneling through the potential barrier between two potential minima. Consequently, farther evolution of this big fluctuation may be studied using the classical general relativity equations for the homogeneous isotropic closed Universe:
\begin{equation} \label{EQ_10}
  \dot{a}^{2} +1=\frac{\tilde{\kappa }}{3} \left(\varepsilon +\lambda \right)a^{2};     \ddot{a}=-\frac{\tilde{\kappa }}{6} \left(\varepsilon +3p-2\lambda \right)a,
\end{equation}
where $a$ is the scale (radius) of the Universe, $\dot{a}$ and $\ddot{a}$ -- its first and second proper time derivatives, $\tilde{\kappa }={\kappa \mathord{\left/ {\vphantom {\kappa  \left(1+2\xi \kappa \phi _{0}^{2} \right)}} \right. \kern-\nulldelimiterspace} \left(1+2\xi \kappa \phi _{0}^{2} \right)} $ is the Einstein gravitational constant, renormalized due to interaction of scalar and gravitation fields (see~\cite{Krive1976}), and parameter $\lambda $ is the vacuum energy density, which is related to the Einstein cosmological constant $\lambda ={\Lambda \mathord{\left/ {\vphantom {\Lambda  \tilde{\kappa }}} \right. \kern-\nulldelimiterspace} \tilde{\kappa }} $.

Notice that the assumption of a rather big initial quantum fluctuation implies the uniqueness of our Universe, as the probability of simultaneous appearance of several such fluctuation is vanishingly small.

With account for a possible time dependence of $\lambda $ equations \eqref{EQ_10} give the energy conservation low:
\begin{equation} \label{EQ_11}
  3\frac{\dot{a}}{a} +\frac{\dot{\varepsilon }+\dot{\lambda }}{\varepsilon +p} =0,
\end{equation}
as well as the expression for the scalar curvature:
\begin{equation} \label{EQ_12}
  R=-\frac{6}{a^{2} } \left(a\ddot{a}+\dot{a}^{2} +1\right)=\tilde{\kappa }\left[(3\nu -1)\varepsilon -4\lambda \right].
\end{equation}

In the absence of matter ($\varepsilon =0$) or for ultrarelativistic matter or equilibrium electromagnetic radiation with equation of state $p=\varepsilon /3$ the scalar curvature of the 4-space equals $R=-4\Lambda \equiv -4\tilde{\kappa }\lambda $. However, for $\nu >1/3$ and $(3\nu -1)\varepsilon >4\lambda $ the scalar curvature is positive.

Contrary to the scenario of ``chaotic inflation''~\cite{Linde1990} where the vacuum energy density is defined as $\lambda =U(\phi )+\dot{\phi }^{2} /2$, due to the constancy of the scalar field amplitude $\phi =\phi _{0} $ and minimal density of the potential energy $U(\phi )=U_{\min }^{\left(1\right)} (\phi _{0} )$ on the phase trajectory $ABC$ we shall assume
\begin{equation} \label{EQ_13}
  \lambda =U_{\min }^{\left(1\right)} (\phi _{0} )=6.75\mu ^{2} \phi _{0}^{2} =const.
\end{equation}

In the capacity of matter filling the early cold Universe we shall consider non-relativistic degenerate Fermi-gas with the equation of state $p=2\varepsilon /3$, consisting of the pairs of fermions and antifermions with masses $m_{F} =m_{AF} $ born in quantum fluctuation. Notice that the finite fermionic mass may be generated due to the interaction between scalar and fermionic fields in accordance with the Higgs mechanism~\cite{Higgs1964,Higgs1964a}. It is assumed that the interaction between fermions is weak enough for the characteristic annihilation time $t_{A} $ of the particles and antiparticles is much greater than the maximal evolution time of the early cold Universe to the point of phase transition (see below).

Thus, assuming $p=2\varepsilon /3$ and $\lambda =const$, in accordance with \eqref{EQ_11} we have:
\begin{equation} \label{EQ_14}
  \varepsilon \left(t\right)=\varepsilon _{0} \cdot \left[{a_{0} \mathord{\left/ {\vphantom {a_{0}  a\left(t\right)}} \right. \kern-\nulldelimiterspace} a\left(t\right)} \right]^{5}.
\end{equation}
Here $\varepsilon _{0} $ and $a_{0} $ are the initial values of the energy density of the matter and radius of the nucleus of the Universe, which satisfy conditions $a_{0} >l_{P} $ and $\varepsilon _{0} \le \varepsilon _{P} $, where $\varepsilon _{P} =M_{P}^{4} $ is the Planck energy density (in the system of units $\hbar =c=1$, when $l_{P} =t_{P} =1/M_{P} $). From \eqref{EQ_12} it follows then:
\begin{equation} \label{EQ_15}
  R(t)=\tilde{\kappa }[\varepsilon (t)-4\lambda ].
\end{equation}

Let us assume that at the initial moment $R\left(0\right)=\tilde{\kappa }(\varepsilon _{0} -4\lambda )>0$ and the curvature value is such that it satisfies the condition $h\left(0\right)\equiv 2\xi R\left(0\right)/\mu ^{2} >0.25$, so that the scalar field potential has a single minimum at the point $\Phi =\phi _{0} $ (see Fig.~\ref{Fig1}).

In the process of the Universe expansion the scalar curvature \eqref{EQ_15} decreases in time with reduction of $\varepsilon \left(t\right)$ in accordance with the power-low dependence \eqref{EQ_14}. In the framework of the proposed model of the nonlinear scalar field this corresponds to the decreasing of the dimensionless parameter of the ``external field'' $h\left(t\right)={2\xi R\left(t\right)\mathord{\left/ {\vphantom {2\xi R\left(t\right) \mu ^{2} }} \right. \kern-\nulldelimiterspace} \mu ^{2} } $. The system is moving then along the phase trajectory $ABC$ (see Fig.~\ref{Fig2}) where the value of the scalar field amplitude $\Phi =\phi _{0} \equiv {\mu \mathord{\left/ {\vphantom {\mu  g}} \right. \kern-\nulldelimiterspace} g} $ and the minimal value of potential $U_{\min }^{\left(1\right)} \left(\phi _{0} \right)=6.75\mu ^{2} \phi _{0}^{2} $ remain constant till the point of phase transition at $h=-2$.

On the assumption of the quantum origin of the Universe it is convenient to introduce dimensionless variables $\tilde{a}\left(\tau \right)={a\left(t\right)\mathord{\left/ {\vphantom {a\left(t\right) l_{P} }} \right. \kern-\nulldelimiterspace} l_{P} } $ and $\tau ={t\mathord{\left/ {\vphantom {t_{P} }} \right. \kern-\nulldelimiterspace} t_{P} } $ (where $t_{P} $ is the Planck time). In this case the scalar curvature \eqref{EQ_15} and dimensionless ``external field'' parameter $h$, with account for the energy conservation \eqref{EQ_14}, are written as:
\begin{equation} \label{EQ_16}
  R\left(\tau \right)=-\Lambda \cdot \left[4-\frac{\varepsilon _{0} }{\lambda } \cdot \left(\frac{\tilde{a}_{0} }{\tilde{a}\left(\tau \right)} \right)^{5} \right];
\end{equation}
\begin{equation} \label{EQ_17}
  h\left(\tau \right)\equiv \frac{2\xi R\left(\tau \right)}{\mu ^{2} } =-\frac{\tilde{\xi }\cdot \beta }{\left(1+\tilde{\xi }\right)} \cdot \left[4-\frac{\tilde{\varepsilon }_{0} }{\beta } \left(\frac{\tilde{a}_{0} }{\tilde{a}\left(\tau \right)} \right)^{5} \right],
\end{equation}
where $\tilde{a}_{0} ={a_{0} \mathord{\left/ {\vphantom {a_{0}  l_{P} }} \right. \kern-\nulldelimiterspace} l_{P} } $, $\tilde{\xi }=2\xi \kappa \phi _{0}^{2} $, $\tilde{\varepsilon }_{0} ={\varepsilon _{0} \mathord{\left/ {\vphantom {\varepsilon _{0}  \mu ^{2} \phi _{0}^{2} }} \right. \kern-\nulldelimiterspace} \mu ^{2} \phi _{0}^{2} } $ and $\beta ={\lambda \mathord{\left/ {\vphantom {\lambda  \mu ^{2} \phi _{0}^{2} }} \right. \kern-\nulldelimiterspace} \mu ^{2} \phi _{0}^{2} } $ are the dimensionless parameters of the present model.

In order to describe the dynamics of the Universe we use the first of the equations \eqref{EQ_10} for the velocity of the expansion (along with the energy conservation low):
\begin{equation} \label{EQ_18}
  \frac{d\tilde{a}}{d\tau } =\left\{b\left[1+\frac{\tilde{\varepsilon }_{0} }{\beta } \cdot \left(\frac{\tilde{a}_{0} }{\tilde{a}\left(\tau \right)} \right)^{5} \right]\cdot \tilde{a}^{2} \left(\tau \right)-1\right\}^{1/2}.
\end{equation}
The quantity $b=\tilde{\kappa }\lambda l_{P}^{2} /3$ is a function of $\tilde{\xi }$ due to the renormalization of the Einstein gravitation constant:
\begin{equation} \label{EQ_19}
  b\left(\tilde{\xi }\right)=\frac{\beta }{3} \frac{\Omega _{P} }{\tilde{\varepsilon }_{P} } \frac{1}{1+\tilde{\xi }},
\end{equation}
where $\Omega _{P} =\kappa \varepsilon _{P} l_{P}^{2} =25.1327...$ is a universal constant, expressed in terms of the world constants, while $\tilde{\varepsilon }_{P} ={\varepsilon _{P} \mathord{\left/ {\vphantom {\varepsilon _{P}  \mu ^{2} \phi _{0}^{2} }} \right. \kern-\nulldelimiterspace} \mu ^{2} \phi _{0}^{2} } $ is an additional dimensionless model parameter, dependent on the parameters of the universal scalar field $\mu $ and $\phi _{0} =\mu /g$. The domain of applicability of the solutions of the classical equation \eqref{EQ_18} should be limited by the conditions $\tilde{a}(\tau )>1$ and $\tau >1$.

The requirement of the real values for the velocity of the early Universe's expansion is equivalent to the requirement of the non-negativeness of the subradical expression in \eqref{EQ_18}, which, with account for \eqref{EQ_19}, is translated into the following conditions:
\begin{equation} \label{EQ_20}
  \tilde{\varepsilon }_{0 \min } \left(\xi \right) =
  \begin{cases}
    \frac{1}{b\left(\tilde{\xi }\right)\tilde{a}_{0}^{2} } -\beta & \text{\hspace{-0.9em} if } \tilde{\xi }\le \frac{5}{9} \beta \frac{\Omega _{P} }{\tilde{\varepsilon }_{P} } \tilde{a}_{0}^{2} ;
    \\
    \frac{2}{3} \beta \left(\frac{5}{3} \beta b\left(\tilde{\xi }\right)\tilde{a}_{0}^{2} \right)^{-\frac{2}{5} } & \text{\hspace{-0.9em} if } \tilde{\xi }>\frac{5}{9} \beta \frac{\Omega _{P} }{\tilde{\varepsilon }_{P} } \tilde{a}_{0}^{2} .
  \end{cases}
\end{equation}

The described scenario assumes that the Universe's expansion may continue only till the time $\tau _{c} $ when parameter $h\left(\tau \right)$ reaches its critical value $h_{c} \equiv h\left(\tau _{c} \right)=-2$ in the point $A$ on the phase trajectory $ABC$ (see. Fig.~\ref{Fig2}) and the dimensionless radius becomes equal to the limit value $\tilde{a}_{c} \equiv \tilde{a}\left(\tau _{c} \right)$.

From \eqref{EQ_17} for $h=-2$ we obtain the ratio ${\tilde{a}_{c} \mathord{\left/ {\vphantom {\tilde{a}_{c}  \tilde{a}_{0} }} \right. \kern-\nulldelimiterspace} \tilde{a}_{0} } $, dependent on the model parameters $\beta $, $\tilde{\varepsilon }_{0} $ and $\tilde{\xi }$:
\begin{equation} \label{EQ_21}
  \frac{\tilde{a}_{c} }{\tilde{a}_{0} } =\left[\frac{\tilde{\xi }}{2\left[\left(2\beta -1\right)\tilde{\xi }-1\right]} \tilde{\varepsilon }_{0} \right]^{{1\mathord{\left/ {\vphantom {1 5}} \right. \kern-\nulldelimiterspace} 5} }.
\end{equation}

The inequality $\tilde{a}_{c} >\tilde{a}_{0} $ necessary in the case of the Universe's expansion leads to the following restriction on the parameters $\tilde{\varepsilon }_{0} $ and $\tilde{\xi }$:
\begin{equation} \label{EQ_22}
  \tilde{\varepsilon }_{0} >2\frac{\left(2\beta -1\right)\tilde{\xi }-1}{\tilde{\xi }}; \quad \tilde{\xi }\ge {1\mathord{\left/ {\vphantom {1 \left(2\beta -1\right)}} \right. \kern-\nulldelimiterspace} \left(2\beta -1\right)}.
\end{equation}

According to \eqref{EQ_21}, when $\tilde{\xi }$ tends to its minimal value $\tilde{\xi }_{\min } =1/(2\beta -1)$ the radius of the Universe in the point of phase transition goes to infinity, $a_{c} \to \infty $.

On the other hand, the initial energy density $\varepsilon _{0} $ of the matter, which was born as the result of the quantum fluctuation of vacuum on the time scale of about $t_{P} $, an not exceed the Planck energy density $\varepsilon _{P} $. Thus the total initial energy of matter $E_{0} =\varepsilon _{0} a_{0}^{3} $ should be bounded from above by the Planck energy $\varepsilon _{P} l_{P}^{3} $, whence we have the inequality:
\begin{equation} \label{EQ_23}
  \tilde{\varepsilon }_{0} \le {\tilde{\varepsilon }_{P} \mathord{\left/ {\vphantom {\tilde{\varepsilon }_{P}  \tilde{a}_{0}^{3} }} \right. \kern-\nulldelimiterspace} \tilde{a}_{0}^{3} }.
\end{equation}

The conditions \eqref{EQ_20}, \eqref{EQ_22} and \eqref{EQ_23}, together with $\tilde{a}_{0} >1$, are the complete set of restrictions, applied to the parameters of the proposed model.

As follows from \eqref{EQ_13}, the dimensionless parameter $\beta $ equals $\beta =6.75$, so the conditions \eqref{EQ_22} may be written as
\begin{equation} \label{EQ_24}
  \tilde{\varepsilon }_{0} >25\frac{(\tilde{\xi }-0.08)}{\tilde{\xi }}; \quad \tilde{\xi }\ge \tilde{\xi }_{\min } =0.08.
\end{equation}

On the other hand, the total potential energy of scalar field, which is released in the first-order phase transition, is determined by the drop of the scalar field potential $\Delta U=6.75\mu ^{2} \phi _{0}^{2} $ and is equal to:
\begin{equation} \label{EQ_25}
  E_{c} =2\pi ^{2} a_{c}^{3} \cdot \Delta U.
\end{equation}

In this case the relation \eqref{EQ_21} and the ratio of the final $E_{c} $ and initial $E_{0} $values of the total energy may be represented as:
\begin{equation} \label{EQ_26}
  \begin{aligned}
    \frac{\tilde{a}_{c} }{\tilde{a}_{0} } &=\left(\frac{0.04\tilde{\xi }}{\tilde{\xi }-0.08} \tilde{\varepsilon }_{0} \right)^{1/5} ; \\ \frac{E_{c} }{E_{0} } &=\frac{\beta }{\tilde{\varepsilon }_{0} } \cdot \left(\frac{\tilde{a}_{c} }{\tilde{a}_{0} } \right)^{3} =\frac{6.75}{\tilde{\varepsilon }_{0}^{{2\mathord{\left/ {\vphantom {2 3}} \right. \kern-\nulldelimiterspace} 3} } } \left\{\frac{0.04\tilde{\xi }}{\tilde{\xi }-0.08} \right\}^{3/5}.
  \end{aligned}
\end{equation}

Thus, in the point of phase transition the maximal radius of the early cold Universe diverge $a_{c} \to \infty $, as well as the total released energy $E_{c} \to \infty $, for all possible initial values of $a_{0} $ and $E_{0} $, if $\tilde{\xi }\to \tilde{\xi }_{\min } =0.08$.

Notice that the value $\tilde{\xi }_{\min } =0.08$ for $\beta =6.75$ corresponds to some minimal value of the interaction constant of scalar and gravitational field $\xi _{\min } =\tilde{\xi }_{\min } /2\kappa \phi _{0}^{2} =0.04/\kappa \phi _{0}^{2} $, which depends on the magnitude of the vacuum average of the scalar field. For example, for the Higgs field with vacuum average $\phi _{H} =\mu _{H} /g_{H} \approx 247$~GeV with good accuracy we have $\kappa \phi _{H}^{2} \approx 10^{-32} $, which gives unrealistically large value $\xi _{\min } \approx 4\cdot 10^{30} $ (cf.~\cite{Zee1979,Smolin1979,Cervantes-Cota1995}) and indicates the impossibility of the direct unification of the standard model of elementary particles with gravitation.

Nevertheless, if we assume that the value of the vacuum average for the fundamental scalar field in the early Universe satisfied condition $\kappa \phi _{0}^{2} \approx 1$, which corresponds to the ratio ${\phi _{0} \mathord{\left/ {\vphantom {\phi _{0}  \phi _{H} }} \right. \kern-\nulldelimiterspace} \phi _{H} } \approx \sqrt{{G_{F} \mathord{\left/ {\vphantom {G_{F}  G}} \right. \kern-\nulldelimiterspace} G} } \approx 10^{16} $, then for the constant $\xi _{\min } $ we obtain a reasonable estimate $\xi _{\min } \approx 0.04$.

In this case the renormalizations of the constant of self-action for the nonlinear scalar field $g^{2} \to g^{2} \cdot (1+\xi \kappa \phi _{0}^{2} )$ and the Einstein gravitational constant $\tilde{\kappa }=\kappa /(1+2\xi \kappa \phi _{0}^{2} )$ (see~\cite{Krive1976}) are about 4\% and 8\% respectively.

\begin{figure}[h]
\includegraphics[width=\columnwidth]{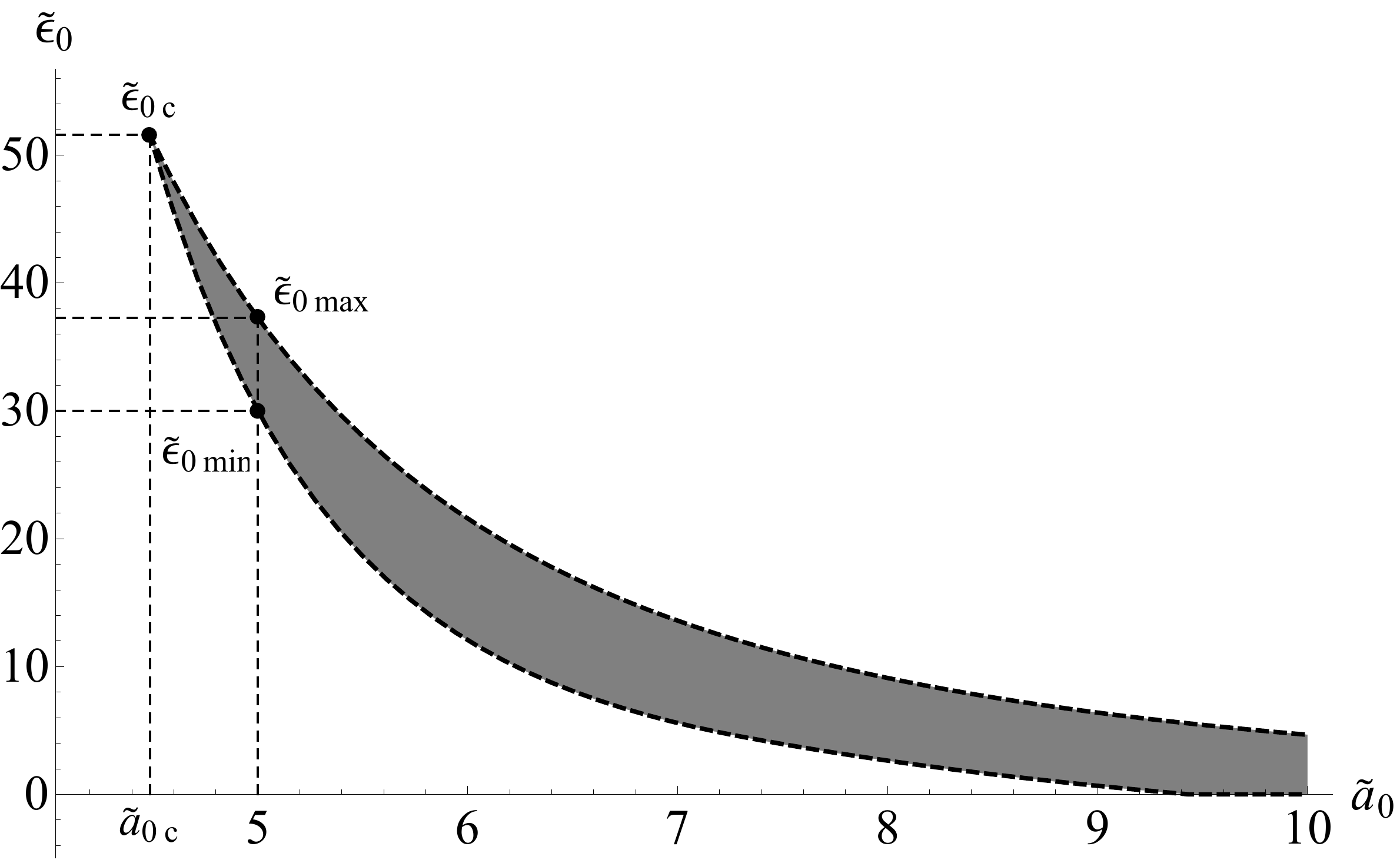}
\caption{\label{Fig3} Two-dimensional domain of existence of solutions for equation \eqref{EQ_18} in the space of dimensionless parameters $\tilde{a}_{0} \equiv {a_{0} / l_{P} } $ and $\tilde{\varepsilon }_{0} \equiv {\varepsilon _{0} / \varepsilon _{P} } $, obtained with account for \eqref{EQ_20}, \eqref{EQ_22} and \eqref{EQ_23}, when dimensionless renormalized constant $\tilde{\xi }$ of scalar field's interaction with gravity equals $\tilde{\xi }=2\xi \kappa \phi _{0}^{2} =0.0801$, which is close to the minimal value $\tilde{\xi }_{\min } =0.08$ for $\beta =6.75$ and $\tilde{\varepsilon }_{P} =4660$.}
\end{figure}

Fig.~\ref{Fig3} represents the domain of existence of solutions for the evolution equations \eqref{EQ_18} in the space of parameters $\tilde{\varepsilon }_{0} $ and $\tilde{a}_{0} $ for $\tilde{\xi }=0.0801$ and $\tilde{\varepsilon }_{P} \equiv \varepsilon _{P} /\mu ^{2} \phi _{0}^{2} =4660$, determined by the restrictions \eqref{EQ_20}, \eqref{EQ_22} and \eqref{EQ_23}. The upper boundary is given by inequality \eqref{EQ_23}, the lower one is defined by the condition $\tilde{\varepsilon }_{0} =\tilde{\varepsilon }_{0 \min } (\tilde{\xi })$ in \eqref{EQ_20}. The calculation is done for $\beta =6.75$ and $\tilde{\varepsilon }_{P} \equiv \varepsilon _{P} /\mu ^{2} \phi _{0}^{2} =4660$. The choice of the value for $\tilde{\varepsilon }_{P} $ corresponds to the condition $\mu /\mu _{H} =\phi _{0} /\phi _{H} $ (see below). For the chosen parameters solutions exist only for sufficiently large initial values of the radius of the Universe $a_{0} >4.5l_{P} $, i.e. only for a rather big quantum fluctuation.

\begin{figure}[h]
\includegraphics[width=\columnwidth]{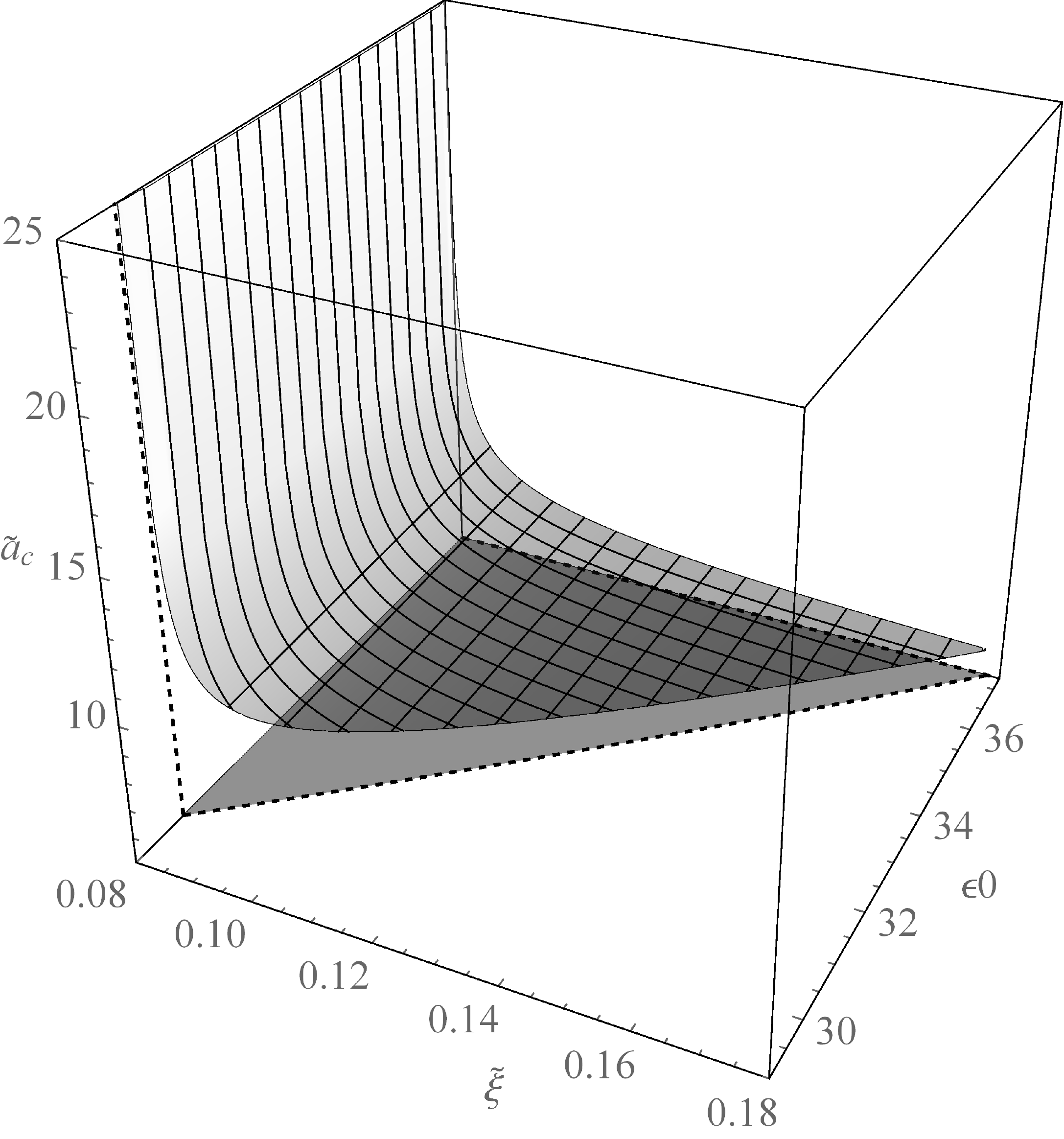}
\caption{\label{Fig4} The dimensionless radius of the Universe $\tilde{a}_{c} $ at the moment of phase transition (see \eqref{EQ_21}) in function of parameters $\tilde{\xi }$ and $\tilde{\varepsilon }_{0} $ for $\tilde{a}_{0} =5$, $\beta =6.75$ and $\tilde{\varepsilon }_{P} =4660$. The dark region in the plane $\tilde{\xi }$~-- $\tilde{\varepsilon }_{0} $ is the domain of allowable values of these parameters, defined by \eqref{EQ_20}, \eqref{EQ_22} and \eqref{EQ_23}.}
\end{figure}

The value of the dimensionless radius of the Universe in the point of phase transition $\tilde{a}_{c} $ is shown in Fig.~\ref{Fig4} in function of the parameters $\tilde{\xi }$ and $\tilde{\varepsilon }_{0} $ for $\tilde{0}_{0} =5$, $\beta =6.75$ and $\tilde{\varepsilon }_{P} =4660$. We can see, that $\tilde{a}_{c} \to \infty $ for $\tilde{\xi }\to 0.08$, in accordance with \eqref{EQ_26}, and significant expansion of the Universe, with $\tilde{a}_{c} \gg \tilde{a}_{0} $, is possible only in a narrow range of the values of $\tilde{\xi }$m when $(\tilde{\xi }-\tilde{\xi }_{\min } )\ll 1$.

\begin{figure}[h]
\includegraphics[width=\columnwidth]{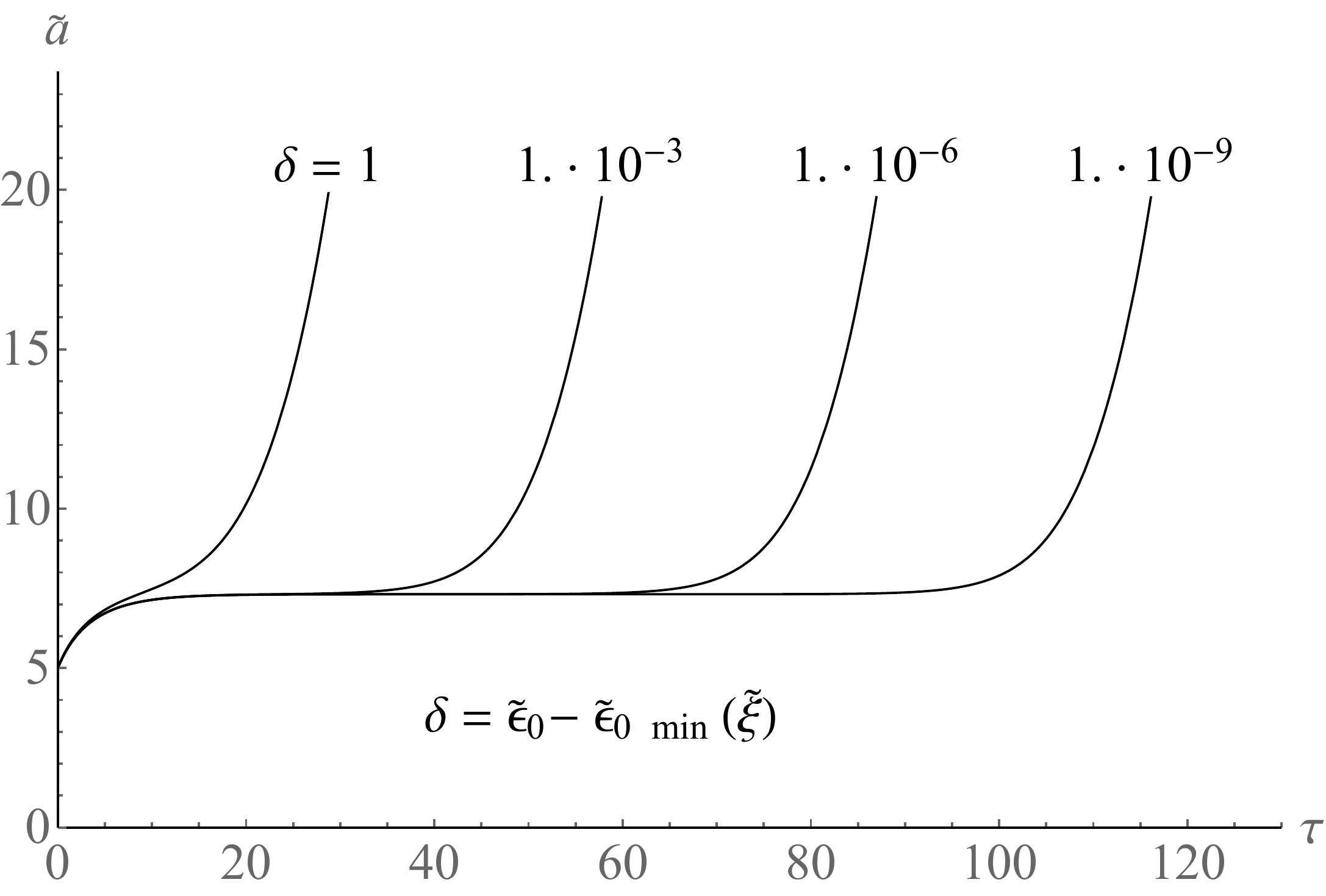}
\caption{\label{Fig5} The dimensionless time $\tau ={t / t_{P} } $ dependencies of the dimensionless radius $\tilde{a}\left(\tau \right)$ of the early cold expanding Universe up to the moment of phase transition $h=h_{c} $. The solutions of equation \eqref{EQ_18} are shown for several values of the parameter $\tilde{\varepsilon }_{0} $ near its minimal value $\tilde{\varepsilon }_{0 \min } $, defined by inequalities \eqref{EQ_20}, for $\tilde{a}_{0} =5$, $\tilde{\xi }=0.0801$, $\beta =6.75$ and $\tilde{\varepsilon }_{P} =4660$.}
\end{figure}

The temporal evolution of the dimensionless radius of the Universe $\tilde{a}(\tau )$ up to the moment of phase transition is illustrated by Fig.~\ref{Fig5}, where solutions of equation \eqref{EQ_18} are shown for several values of the parameter $\tilde{\varepsilon }_{0} $ in the vicinity of its minimal value $\tilde{\varepsilon }_{0 \min } (\xi )$, determined by the relations \eqref{EQ_20}. In this case the maximal radius of the Universe, which is restricted by the phase transition at the moment $t=t_{c} $, is almost independent of $\tilde{\varepsilon }_{0} $ and equals $a_{c} \approx 20l_{P} $, while the time of the Universe's evolution changes in a wide range $25t_{P} <t<120t_{P} $ owing to the grows of the ``plateau''. In all cases though, at the later stage, when $(\tilde{a}_{c} /\tilde{a}_{0} )^{5} \gg 1$, the Universe expands in accordance with exponential low, typical for inflationary solutions. It should be emphasized the in this case, similarly to the de Sitter model, inflation is driven by the constant vacuum energy density $\lambda =const$, contrary to the scenario of ``chaotic inflation''~\cite{Linde1982,Linde1983,Linde1983a,Linde1990}, where the expansion of the Universe occurs on the background of the diminishing energy density of the scalar field.

\begin{figure}[h]
  \subfloat[]{\label{Fig6a} \includegraphics[width=.98\columnwidth]{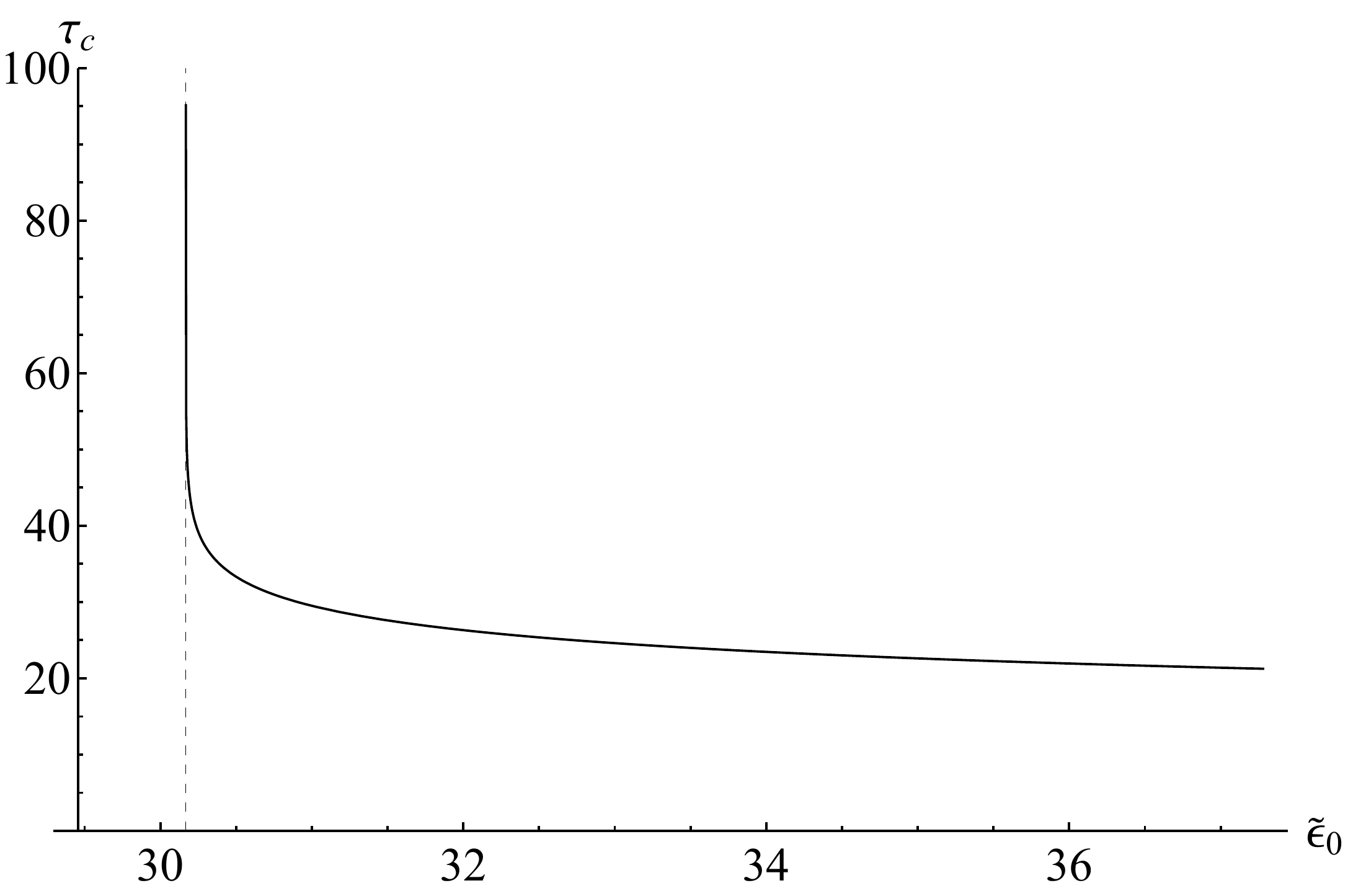}}\\
  \begin{flushright}
  \subfloat[]{\label{Fig6b} \includegraphics[width=.97\columnwidth]{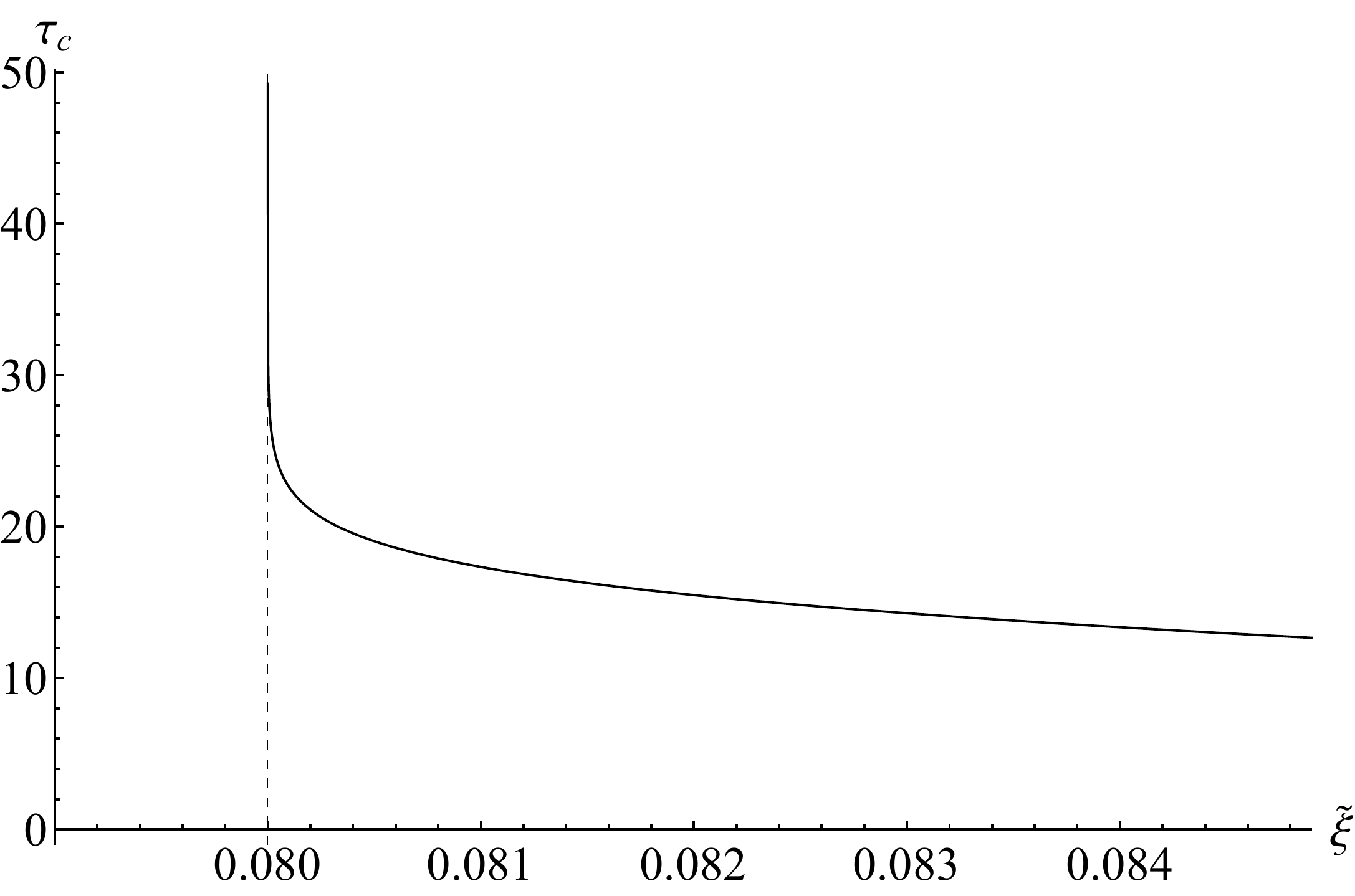}}
  \end{flushright}
  \caption[]{\label{Fig6} The duration of the early Universe's expansion up to the moment of phase transition in function of $\tilde{\varepsilon }_{0} $ in the vicinity of the limiting value $\tilde{\varepsilon }_{0 \min } \approx 30.17$ for $\tilde{\xi }=0.0801$ \subref{Fig6a} and in function of $\tilde{\xi }$ near the minimal value $\tilde{\xi }_{\min } =0.08$ for $\tilde{\varepsilon }_{0} =35$ \subref{Fig6b}. Both curves are plotted with $\tilde{a}_{0} =5$, $\beta =6.75$ and $\tilde{\varepsilon }_{P} =4660$.}
\end{figure}

The dimensionless duration of the early Universe expansion $\tau _{c} \equiv t_{c} /t_{P} $ is shown in Fig.~\ref{Fig6} in functions of $\tilde{\varepsilon }_{0} $ and $\tilde{\xi }$. The unlimited grows of the time $\tau _{c} \to \infty $ for $\tilde{\varepsilon }_{0} \to \tilde{\varepsilon }_{0 \min } $ (Fig.~\ref{Fig6a}) is due the fact that for $\tilde{\varepsilon }_{0} =\tilde{\varepsilon }_{0 \min } (\tilde{\xi })$ the minimal value of the expansion velocity $\dot{a}$ becomes zero with simultaneous vanishing of acceleration $\ddot{a}$ and all higher time derivatives of $a$. As $\tilde{\varepsilon }_{0} $ approaches $\tilde{\varepsilon }_{0 \min } $($\tilde{\xi }$) on the lower boundary of the shaded region in Fig.~\ref{Fig3} the plateau on the curve $\tilde{a}(\tau )$ (see Fig.~\ref{Fig5}) may stretch to infinity, provided $\tilde{a}_{c} $ is larger than the value
\begin{equation} \label{EQ_27}
  \tilde{a}_{c\min } =\tilde{a}_{0} \cdot (3\tilde{\varepsilon }_{0} /2\beta )^{1/5},
\end{equation}
which corresponds to the minimal value of $\dot{a}$ according to \eqref{EQ_18}.

\section{\label{Sec_Model} Model parameters' estimates and evolution of the Universe after the first-order phase transition}

Let us make estimates of the scalar field parameters in the framework of the proposed model of the early cold Universe. As was shown in section~\ref{Sec_Evolution}, a physically reasonable assessment of the dimensionless constant of interaction of scalar and gravitational fields $\xi \approx 0.04$ may be achieved with assumption $\kappa \phi _{0}^{2} \approx 1$, which corresponds to a rather large ratio $\phi _{0} /\phi _{H} \approx 10^{16} $ of the vacuum averages of the fundamental scalar field $\phi _{0} =\mu /g$ and the Higgs field $\phi _{H} =\mu _{H} /g_{H} $.

On the other hand, the choice of the big value for the dimensionless parameter $\tilde{\varepsilon }_{P} \equiv \varepsilon _{P} /\mu ^{2} \phi _{0}^{2} =4660$, with $\varepsilon _{P} =M_{P}^{4} $, corresponds to the ratio $\mu \phi _{0} /\mu _{H} \phi _{H} \approx 10^{32} $. Both these choices may by consistent only if $\mu /\mu _{H} \approx 10^{16} $ and $g/g_{H} \approx 1$, which also gives $\mu \approx 0.1M_{P} $ and $\phi _{0} =\mu /g\approx 0.274M_{P} $ for $g\approx g_{H} \approx 0.364$. The ratio of the potential energy density of the scalar field to the Planck energy density then equals $\Delta U/\varepsilon _{P} =6.75\mu ^{2} \phi _{0}^{2} /M_{P}^{4} \approx 0.00145$.

Accordingly, the fundamental scalar field in the early cold Universe prior to Big Band could have the potential energy density $\Delta U=6.75\mu ^{4} /g^{2} $, which is 64 orders of magnitude larger than that value for the Higgs field, once more emphasizing the lack of usability of the latter in the inflation theories of the early Universe.

In the model of ``chaotic inflation''~\cite{Linde1990}, with the restriction on the scalar field potential $V(\phi )\le M_{P}^{4} $, for the size of the ``inflated'' Universe to exceed by many orders of magnitude the size of observable Universe it is necessary to have large initial amplitude of the scalar field $\phi \gg M_{P} $ and anomalously small values of either the effective mass of the scalar field $m\ll M_{P} $ for the quadratic potential $V(\phi )=m^{2} \phi ^{2} /2$, or the nonlinearity coefficient $\gamma \ll 1$ for the potential $V(\phi )=\gamma \phi ^{4} $. On the other hand, it should be stressed that in the framework of the proposed model with $\xi \to \xi _{\min } =0.04$ the Universe's inflation to arbitrary large sizes in the point of the phase transition is possible for $\phi _{0} <M_{P} $ and $\Delta U\ll \varepsilon _{P} $.

Accordingly, the presently proposed scenario of evolution of the early Universe towards the point of phase transition with unrestricted ``inflation'' for $\xi \to \xi _{\min } $, may be called ``hyperinflation''.

Let us estimate the total energy of the scalar field $E_{c} $, freed in the first-order phase transition, with account for the parameters' values $a_{c} \approx 20l_{P} $ and $\Delta U\approx 1.45\cdot 10^{-3} \cdot M_{P}^{4} $, obtained above:
\begin{equation}  \label{EQ_28}
  E_{c} =2\pi ^{2} a_{c}^{3} \cdot \Delta U\approx 2.3\cdot 10^{2} \cdot M_{P} \approx 2.76\cdot 10^{21} ~\text{GeV}.
\end{equation}

The discharge of such enormous energy during phase transition should lead to the birth of a huge amount of particle-antiparticle pairs of various kinds, and to the rapid heating of the matter to high temperature of the order of the Planck one $T_{P} \approx M_{P} /k_{B} \approx 1.2\cdot 10^{32} $~K ($k_{B} $ is the Boltzmann constant), thus causing the ``Big Bang'' starting the hot phase of our Universe.

On the other hand, potential \eqref{EQ_7} for $h=-2$ with zero minimum in the point $x=-2$ (see Fig.~\ref{Fig1}) may be rewritten with shifted field amplitude $y=x+2$:
\begin{equation} \label{EQ_29}
  V(y)=\frac{9}{2} y^{2} -2y^{3} +\frac{y^{4} }{4}.
\end{equation}
One may assume it to be an analog of various potentials, considered in the models of ``chaotic inflation''~\cite{Linde1990}. In this case the frequency of oscillations of the scalar field amplitude near the minimum at $y=0$ equals $\omega =3\mu \approx 0.3M_{P} $.

In this way the scenario of evolution of the early cold Universe, considered here, may be connected to a subsequent process of evolution described by the model of ``chaotic inflation''~\cite{Linde1990}, which allows to solve many problems of cosmology. Nevertheless, the problems of dynamics of scalar field with potential \eqref{EQ_29} and farther evolution of the Universe with its heating after the first-order phase transition go beyond the scope of the present paper.

\section{Conclusions}

We have proposed a scenario of evolution of the early cold Universe which appears as the result of a rather big quantum fluctuation of vacuum with consequent first-order phase transition, driven by an ``external field'' parameter, which is proportional to the time-dependent scalar curvature $R(t)$. It is assumed that this phase transition occurs in the expanding Universe due to the interaction of the fundamental nonlinear scalar field with gravitational field, on the one hand, and also because of the presence of matter with equation of state $p=\nu \varepsilon $ with $\nu >{1\mathord{\left/ {\vphantom {1 3}} \right. \kern-\nulldelimiterspace} 3} $, on the other hand. The solutions of the nonlinear general relativity equations with finite energy density of vacuum may exist only for a rather large initial radius $a_{0} \ge 5l_{P} $ of the incipient Universe, when the quantum effects may be considered small, justifying the applicability of the classical relativity equations. The probability of such a big fluctuation is by itself rather low, while simultaneous appearance of spatially close multiple fluctuations is most improbable, which implies uniqueness of the Universe, developed in accordance with the described scenario.

We have obtained estimates for various parameters of the model: for $\mu $ and $g$, and also the vacuum average $\phi _{0} =\mu /g$ of the fundamental nonlinear scalar field; for the vacuum energy density, which is determined by the density of the potential energy of the scalar field $\lambda =\Delta U$; for the constant of interaction of the scalar and gravitational fields $\xi $; for the total energy of the scalar field $E_{c} =\Delta U\cdot \upsilon _{c} $, which is released during the first-order phase transition in the whole volume of the closed Universe $\upsilon _{c} =2\pi ^{2} a_{c}^{3} $, where $a_{c} $ is the maximal radius of the early Universe in the point of phase transition. We have shown that when the parameter $\xi $ approaches some limit value $\xi _{\min } $ (dependent on the value of the vacuum average of the scalar field $\phi _{0} $) the radius $a_{c} $ and, consequently, energy $E_{c} $ tend to infinity, which may be called a ``hyperinflation'' regime of evolution of the early Universe.

\begin{acknowledgments}
In conclusion, we would like to express our gratitude to D.S.~Gorbunov, G.M.~Zinoviev, A.I.~Zhuk, I.V.~Krive, V.V.~Lebedev, V.A.~Rubakov and S.M.~Ryabchenko for enlightening discussions and useful criticism.
\end{acknowledgments}

\hyphenation{Cosmology}
\bibliography{Bang_new_translation}

\end{document}